\documentclass[11pt]{article}

\usepackage{graphicx,amsmath,amsmath,amsfonts,amssymb,dcolumn,amsthm,euscript}

\begin{document}

%




\date{}
\title{\textbf{Renormalizability of pure $\mathcal{N}=1$ Super Yang-Mills  in the Wess-Zumino gauge in the presence of the local composite operators $A^{2}$ and $\bar{\lambda}\lambda$
}}


 \author{\textbf{M.~A.~L.~Capri}\thanks{caprimarcio@gmail.com}$\,\,\,\,^{a}$\,\,,
  \textbf{S.~P.~Sorella}\thanks{silvio.sorella@gmail.com}$\,\,\,\,^{a}$\,\,,
 \textbf{R.~C.~Terin}\thanks{rodrigoterin3003@gmail.com}$\,\,\,\,^{a}$\,\,,
 \textbf{H.~C.~Toledo}\thanks{henriqcouto@gmail.com}$\,\,\,\,^{a}$
 \\[2mm]
 {\small \textnormal{ \it $^{a}$  UERJ -- Universidade do Estado do Rio de Janeiro,}}\\
 {\small \textnormal{ \it Instituto de F\'{\i}sica, Departamento de F\'{\i}sica Te\'{o}rica,}}\\
 {\small \textnormal{ \it Rua S\~{a}o Francisco Xavier 524, 20550-013 Maracan\~{a}, Rio de Janeiro, RJ, Brasil}}
  \\[2mm]
} 	

\maketitle 

\begin{abstract}
The $\mathcal{N}=1$ Super Yang-Mills theory in the presence of the local composite operator $A^2$  is analyzed in the Wess-Zumino gauge  by employing the Landau gauge fixing condition. Due to the superymmetric structure of the theory, two more composite operators, $A_\mu \gamma_\mu \lambda$ and $\bar{\lambda}\lambda$, related to the susy variations of $A^2$
are also introduced. A BRST invariant action containing all these operators is obtained. An all order proof of the multiplicative renormalizability of the resulting theory is then provided by means of the algebraic renormalization setup. Though, due to the non-linear realization of the supersymmetry in the Wess-Zumino gauge, the renormalization factor of the gauge field turns out to be different from that of the gluino.

\end{abstract}

\section{Introduction}
\hspace{0.5 cm} The supersymmetric $\mathcal{N}=1$ gauge theories  exhibit  remarkable properties, both at perturbative as well as at the non-perturbative levels, see, for example,  \cite{Veneziano:1982ah,Amati:1988ft,Gates:1983nr,Shifman:1978bx} and references therein.\\\\In the last decades we have witnessed a growing interest concerning the dimension two operator $A^2$ and its  possible role in the understanding of non-perturbative aspects of Yang-Mills theories. For instance, the existence of the non-perturbative condensate $\langle A^2 \rangle$ in the Landau gauge has been largely exploited from theoretical, phenomenological and numerical viewpoints, see  \cite{Cornwall:1981zr,Greensite:1985vq,Stingl:1985hx,Lavelle:1988eg,Gubarev:2000eu,Gubarev:2000nz,
Boucaud:2002fx,Kondo:2001nq,Kondo:2001tm,
Verschelde:2001ia,Dudal:2002pq,Dudal:2003vv,Browne:2003uv,Dudal:2003by,Dudal:2004rx,Browne:2004mk,
Gracey:2004bk,Li:2004te,Boucaud:2001st,Boucaud:2002nc, Boucaud:2005rm, RuizArriola:2004en,Boucaud:2005xn,Boucaud:2008gn,Pene:2011kg,Dudal:2005na,Furui:2005bu,
Gubarev:2005it,Slavnov:2005av,Suzuki:2004dw,Suzuki:2004uz,Chernodub:2005gz,
Boucaud:2010gr,Blossier:2010ky,Dudal:2010tf,Boucaud:2011eh,Blossier:2011tf,Blossier:2013te,Fiorentini:2016rwx} for a general overview. \\\\For instance, in \cite{Gubarev:2000eu}, the idea that the dimension two condensate $\langle A^2 \rangle$  enters the operator product expansion (OPE) of the gluon propagator has been put forward. Moreover, a combined OPE and lattice analysis has shown that this condensate can account for the $1/Q^{2}$ corrections which have been reported \cite{Boucaud:2001st,Boucaud:2002nc,Boucaud:2005rm,RuizArriola:2004en,Boucaud:2005xn,
Boucaud:2008gn,Pene:2011kg,Furui:2005bu,Boucaud:2010gr} in the running of the coupling constant as well as  in the two-point gluon correlation function.\\\\Furthermore, an effective potential for $\langle A_{\mu}^{a}A_{\mu}^{a}\rangle$ in Landau gauge has been obtained and evaluated in analytic form at two loops \cite{Verschelde:2001ia,Dudal:2003vv,Browne:2003uv,Browne:2004mk,
Gracey:2004bk}, showing that a non-vanishing value of $\langle A_{\mu}^{a}A_{\mu}^{a}\rangle$ is favoured as it lowers the vacuum energy. Consequently, a dynamical gluon mass is generated. \\\\Let us also  remind here that, in the Landau gauge, the local operator $A_{\mu}^{a}A_{\mu}^{a}$ is BRST-invariant on-shell, a property which has allowed for an all-orders proof of its multiplicative renormalizability \cite{Dudal:2002pq}. Interestingly, its anomalous dimension is not an independent parameter of the theory, being given by a combination of the gauge $\beta$-function and of the anomalous dimension of the gauge field $A_{\mu}^{a}$ \cite{Dudal:2002pq}, namely
\begin{eqnarray}
\gamma_{A^{2}}\Big|_{Landau} &=& \Bigg(\frac{\beta(a)}{a} + \gamma_{A}^{Landau}(a)\Bigg),\,\,\,\,\, a = \frac{g^{2}}{16\pi^{2}},    \label{nrAA}
\end{eqnarray}
where $(\beta(a), \gamma_{A}^{Landau}(a))$ denote, respectively, the $\beta$-function and the anomalous dimension of the gauge field $A_{\mu}^{a}$ in the Landau gauge. This relation was conjectured and verified up to three-loop order in \cite{Gracey:2004bk}. Its validity at all orders of the loop expansion can be found in \cite{Dudal:2002pq}. \\\\The aim of this work is that of extending the previous analysis \cite{Dudal:2002pq} about the operator $A^2$ to the case of $\mathcal{N}=1$ Super Yang-Mills in Euclidean space-time. Here, next to the operator $A^2$ we are naturally led to take into account other composite operators related to the superysmmetric variations of $A^2$, namely $A_\mu \gamma_\mu \lambda$ as well as the gluino operator $ \bar{\lambda}\lambda$. As one can easily figure out, these operators arise when taking the variations  $\delta_{\alpha} A^2$ and $\delta^{\alpha} \delta_{\alpha} A^2$, where $\delta_{\alpha}$, $\alpha=1,2,3,4$, stand for the susy generators. We shall thus analyse the all order renormalizability of $\mathcal{N}=1$ Super Yang-Mills in the presence of the aforementioned operators. For such a purpose  the Wess-Zumino gauge will be employed, where the number of field components is minimum. Also, we shall make use of the Landau gauge-fixing condition, properly adapted to the supersymmetric case.  \\\\In order to achieve an all order proof,  the algebraic renormalization setup \cite{Piguet:1995er} will be exploited. As we shall see, despite the large number of fields and composite operators present in the theory, a very limited number  of renormalization factors will be needed to ensure renormalizability.  The present study can be seen as a sequence of previous works \cite{White:1992ai,Maggiore:1994dw,Maggiore:1994xw,Maggiore:1995gr,
Maggiore:1996gg,Ulker:2001rc,Capri:2014jqa}, where the renormalization of supersymmetric gauge theories through the BRST cohomology and of the algebraic procedure has been pursued. \\\\ 
Despite the large quantity of results that has been obtained so far about dimension two operators, it's worth to emphasise that many aspects related to them still deserve a better understanding. This is the case of the potential role that the condensate  $\langle AA\rangle$  might have in $\mathcal{N}=1$ Super Yang-Mills theories, next to the well known condensate  $ \langle \bar{\lambda}\lambda \rangle $ \cite{Veneziano:1982ah}.  In this sense, the present work can be seen as a first step towards such a goal. \\\\The paper is organized as follows: in Section 2, we provide a brief review of the main aspects of the quantization of  Euclidean $\mathcal N=1$ super Yang-Mills  in the Wess-Zumino gauge. In Section 3, we discuss the introduction of the local operators $A^2$,  $A_\mu \gamma_\mu \lambda$ and $\bar{\lambda}^{\alpha}\lambda_{\alpha}$. In Section 3, we determine the Ward identities fulfilled by the resulting quantized action. In Section 4, we construct the most general counterterm and we establish the renormalization factors for all fields, external sources and parameters. In Section 5, we present our conclusion. The final Appendices contain all conventions and notations.

\section{ Quantization of $\mathcal N=1$ Euclidean Super Yang-Mills in the Wess-Zumino gauge}

\hspace{0.5 cm} As we have mentioned in the Introduction, we shall employ  the Wess-Zumino gauge because of the minimum number of field components exhibited. However, there is a little disadvantage: the supersymmetry algebra is realized in a non-linear way. As a consequence, the algebra of the susy generators; $\delta_{\alpha}$, $\alpha=1,2,3,4$, does not close on the translations. Rather, we have 
\begin{equation} 
\{ \delta_\alpha, \delta_\beta \} = (\gamma_{\mu})_{\alpha\beta} \partial_{\mu} \; + \;\; ({\rm gauge \; transf.}) \; + \;\; ({\rm field \; eqs.}) \;. \label{nlalg}  
\end{equation}
As shown in \cite{White:1992ai,Maggiore:1994dw,Maggiore:1994xw,Maggiore:1995gr,Maggiore:1996gg,Ulker:2001rc}, the most efficient way to handle this kind of algebra, eq.\eqref{nlalg}, is to  construct a generalized BRST operator $Q$ which encodes both susy and gauge transformations, {\it i.e.}
\begin{equation} 
Q=s+\epsilon^{\alpha}\delta_{\alpha} \;, \label{Q}
\end{equation}
where $s$ is the standard BRST operator for the gauge transformations and $\epsilon^{\alpha}$ is a constant susy parameter, namely a Majorana constant spinor (see Appendix \ref{notations}), which carries ghost number +1. It can also be interpreted as a constant ghost for the susy generators. The  operator $Q$ enjoys the following property 
\begin{equation} 
Q^{2} = \epsilon^{\alpha}(\gamma_{\mu})_{\alpha\beta}\bar{\epsilon}^{\beta}\partial_{\mu} \;, \label{Q2}
\end{equation}
which will allow to  quantize the theory in a way in which supersymmetry is manifestly preserved by the gauge  fixing procedure. In particular, from eq.\eqref{Q2}, it follows that the operator $Q$ is nilpotent when acting on space-time integrated quantities. \\\\Let us thus start by giving the classical action of $\mathcal N=1$ Super Yang--Mills theory in the Euclidean space, namely
\begin{equation}
\label{SYM}
\Sigma_\text{SYM} = \int d^{4}x \left[ \frac{1}{4}F^{a}_{\mu \nu}F^{a}_{\mu\nu} 
+ \frac{1}{2} \bar{\lambda}^{a\alpha} (\gamma_{\mu})_{\alpha\beta} D^{ab}_{\mu}\lambda^{b\beta}
+ \frac{1}{2}\mathfrak{D}^a\mathfrak{D}^a\right]\;,  
\end{equation}
where $D^{ab}_{\mu}= (\delta^{ab} \partial_\mu - g f^{abc}A^c_\mu)$ is the covariant derivative in  the adjoint representation of the $SU(N)$ symmetry group, $\lambda^{a\alpha}$ is a Majorana spinor\footnote{Notice that $\bar{\lambda}$ is not independent from $\lambda$ in the Majorana representation, see Appendix \ref{notations} for details.},   $\mathfrak{D}^a$ is an auxiliary field\footnote{Further considerations about the auxiliary field $\mathfrak{D}^a$ will be given at the end of this section.} and
\begin{equation}
F^{a}_{\mu\nu} = \partial_{\mu}A^{a}_{\nu} - \partial_{\nu}A^{a}_{\mu} + gf^{abc}A^{b}_{\mu}A^{c}_{\nu}\;.
\end{equation}
According to the construction of the generalized BRST operator $Q$, eq.\eqref{Q}, its action on  each field is given by  
\begin{eqnarray}
\label{susytransf}
&&
QA^{a}_{\mu} = - D^{ab}_{\mu}c^{b} 
+\bar{\epsilon}^\alpha(\gamma_\mu)_{\alpha\beta}\lambda^{a\beta}\;,\nonumber \\
&&
Q\lambda^{a\alpha} = gf^{abc}c^{b}\lambda^{c\alpha}
- \frac{1}{2}(\sigma_{\mu\nu})^{\alpha\beta}\epsilon_{\beta} F_{\mu\nu}^{a}
+ (\gamma_{5})^{\alpha\beta}\epsilon_{\beta} \mathfrak{D}^a\;, \nonumber \\
&&
Q\mathfrak{D}^a = gf^{abc}c^{b}\mathfrak{D}^c 
+ \bar{\epsilon}^{\alpha}(\gamma_{\mu})_{\alpha\beta}(\gamma_{5})^{\beta\eta}D_{\mu}^{ab}\lambda^{b}_{\eta} \;, \\
&&
Qc^{a} = \frac{1}{2}gf^{abc}c^{b}c^{c} 
- \bar{\epsilon}^{\alpha}(\gamma_{\mu})_{\alpha\beta}\epsilon^{\beta} A^{a}_{\mu}\;, \nonumber \\
&&
Q\bar{c}^{a} = b^{a}\;, \nonumber \\
&&
Qb^{a} = \nabla\bar{c}^{a} \;,\nonumber\\
&& 
Q^2= \nabla  \nonumber \;, 
\end{eqnarray}
where we have introduced the translation operator 
\begin{equation}
\label{top}
\nabla := \bar{\epsilon}^{\alpha}(\gamma_{\mu})_{\alpha\beta}\epsilon^{\beta} \partial_{\mu}\,.
\end{equation}
The fields $({\bar c}^a, c^a)$ denote the Faddeev-Popov ghosts, while $b^a$ is a Langrange multiplier,  necessary in order to implement the  Landau gauge fixing condition, $\partial_\mu A^a_\mu=0$. \\\\It turns out that   the action \eqref{SYM} is left invariant by the previous transformations \eqref{susytransf}, {\it i.e.}  
\begin{equation} 
Q \Sigma_\text{SYM} = 0 \;. \label{invQ}
\end{equation}
In order to quantize this theory, we have to introduce the gauge fixing term. Following the BRST procedure, this is done by adding to the action \eqref{SYM}  an exact $Q$-term.  Adopting the Landau gauge condition, $\partial_\mu A^a_\mu=0$, for the gauge fixing term  we write  
\begin{equation}
\Sigma_\text{gf} = Q\int d^{4}x (\bar{c}^{a}\partial_{\mu}A^{a}_{\mu})\;,  \label{gfx}
\end{equation}
which, according to eqs.\eqref{susytransf}, reads
\begin{equation}
\Sigma_\text{gf} = \int d^{4}x \left[ \bar{c}^{a}\partial_{\mu}D^{ab}_{\mu}c^{b} 
+ b^{a}\partial_{\mu}A^{a}_{\mu} 
- \bar{c}^{a}\bar{\epsilon}^{\alpha}(\gamma_{\mu})_{\alpha\beta}\partial_{\mu}\lambda^{a\beta} \right]\;.  \label{gfx1}
\end{equation}
Therefore, for the super Yang-Mills action quantized in the Landau we have
\begin{eqnarray}
\label{act1}
\Sigma &=& \Sigma_{SYM} + \Sigma_\text{gf} \nonumber \\
&=&
\int d^{4}x \left\{\frac{1}{4}F^{a}_{\mu \nu}F^{a}_{\mu\nu} 
+ \frac{1}{2}\bar{\lambda}^{a\alpha}(\gamma_{\mu})_{\alpha \beta}D^{ab}_{\mu} \lambda^{b\beta}
+ \frac{1}{2}\mathfrak{D}^{2} \right. \nonumber \\
&&
\left.
+ b^{a}\partial_{\mu}A^{a}_{\mu}
+\bar{c}^{a}\left[\partial_{\mu}D^{ab}_{\mu}c^{b} - \bar{\epsilon}^{\alpha}(\gamma_{\mu})_{\alpha \beta}\partial_{\mu}\lambda^{a\beta} \right]\right\}\,.
\label{gfaction}
\end{eqnarray}
From equations \eqref{susytransf}, \eqref{invQ}, \eqref{gfx}, it immediately follows that  
\begin{equation}
Q \Sigma = 0 \;, \label{invS}
\end{equation}
from which one sees that the gauge-fixing procedure has been  implemented in a BRST invariant way. It is worth pointing out that the generalized BRST operator $Q$ encodes  both susy and gauge transformations. As such, equation \eqref{gfx1} represents the supersymmetric generalization of the standard Landau gauge, a fact that can be easily understood through the presence of the additional term $\bar{c}^{a}\bar{\epsilon}^{\alpha}(\gamma_{\mu})_{\alpha\beta}\partial_{\mu}\lambda^{a\beta}$, which contains the supersymmetric ghost, $\bar{\epsilon}^{\alpha}$, as well as the gluino field $\lambda^{a\beta}$. \\\\Before  ending this section, let us spend a few words on the role played by the auxiliary field $\mathfrak D^{a}$ in the starting action 
\eqref{SYM}.  As we shall see later, this field is needed in order to write down a suitable set of Ward identities, which will be employed for the algebraic renormalization procedure \cite{Piguet:1995er}. However, one can observe that $\mathfrak D^{a}$ appears only  at the quadratic level in expression \eqref{SYM}.  As such, we have here two equivalent options: the first one is to keep the $\mathfrak D^{a}$ field in the starting action \eqref{SYM} as well as in the generalized BRST $Q$-transformations, eq.\eqref{susytransf}. In this case, the BRST operator  $Q$ enjoys the property
\begin{eqnarray}
Q^2= \nabla \,,
\label{Q5}
\end{eqnarray}
from which the Slavnov-Taylor identity can be cosntructed in the standard way  \cite{Piguet:1995er}.  The second option would be that of not including  the field $\mathfrak D^{a}$  from the beginning as, for instance, done in   \cite{Maggiore:1994dw,Maggiore:1994xw,Maggiore:1995gr}. This means that the $\mathfrak D^{a}$ field is not present in the starting action as well as in  the  $Q$-transformations. However, in this second case, the operator $Q$ does not enjoy property \eqref{Q5}. Rather, we would have
\begin{equation}
Q^2= \nabla + {\rm eqs. \; of\; motion} \,,
\label{Q20}
\end{equation}
showing that the algebra now closes only on-shell, {\it i.e.} modulo the equations of motion. In this case,  to establish the Slavnov-Taylor identity would require the introduction of terms quadratic in the external BRST sources, see \cite{Maggiore:1994dw,Maggiore:1994xw,Maggiore:1995gr}. These quadratic terms in the external BRST sources play exactly the same role of the term  $\mathfrak D^{a}\mathfrak D^{a}$. At the end, both options will give rise to the same result. In the present work we keep the first option, {\it i.e.} we retain  the   bilinear term $\mathfrak D^{a}\mathfrak D^{a}$ in the starting action.

\section{ Introducing the local composite operators $A_{\mu}A_{\mu}$ and $\bar{\lambda}^{\alpha}\lambda_{\alpha}$}
\label{sec2}


Having introduced the gauge-fixing term, eq.\eqref{gfx1}, let us proceed to establish the Ward identities of the model. Following the algebraic renormalization setup \cite{Piguet:1995er}, we first need to introduce a suitable set of external BRST sources in order to define the non-linear transformations of the fields appearing in eqs.\eqref{susytransf}. More precisely, we need to introduce external sources coupled to the composite operators $QA^a_\mu$, $Q\lambda^{a\beta}$, $Q\mathfrak{D}^a$ and $Qc^a$.  Due to property \eqref{Q5}, this task can be achieved by introducing the following $Q$-doublets of external sources \cite{Piguet:1995er}, namely 
\begin{equation}
\left\{\begin{matrix}QK^{a}_{\mu}=\Omega^{a}_{\mu}\phantom{\Bigl|}\cr
Q\Omega^{a}_{\mu}=\nabla K^{a}_{\mu}\phantom{\Bigl|}\end{matrix}\right.\,,\qquad
\left\{\begin{matrix}QL^{a}=\Lambda^{a}\phantom{\Bigl|}\cr
Q\Lambda^{a}=\nabla L^{a}\phantom{\Bigl|}\end{matrix}\right.\,,\qquad
\left\{\begin{matrix}QT^{a}= J^{a}\phantom{\Bigl|}\cr
QJ^{a}=\nabla T^{a}\phantom{\Bigl|}\end{matrix}\right.\,,\qquad
\left\{\begin{matrix}QY^{a\alpha}=X^{a\alpha}\phantom{\Bigl|}\cr
QX^{a\alpha}=\nabla Y^{a\alpha}\phantom{\Bigl|}\end{matrix}\right.\,.
\end{equation}
Accordingly, the corresponding $Q$ invariant external source term to be added to the action \eqref{act1}  is given by 
\begin{equation}
\Sigma_\text{ext} = Q \int d^{4}x \left( -K^{a}_{\mu} A^{a}_{\mu} + L^{a}c^{a} - T^{a} \mathfrak{D}^a  + Y^{a\alpha}\lambda^{a}_{\alpha} \right)\;,  \label{qexact}
\end{equation}
leading to the following $Q$ invariant action $\Sigma_{0}$
\begin{equation}
\Sigma_{0} = \Sigma_{SYM} + \Sigma_\text{gf} + \Sigma_\text{ext}   \;. \label{complact}
\end{equation}
\begin{equation}
Q \Sigma_{0} = 0 \;. 
\end{equation}
Explicitly, we have 
\begin{eqnarray}
\label{fullact}
\Sigma_{0} &=& \int d^{4}x \biggl\{ \frac{1}{4}F^{a}_{\mu \nu}F^{a}_{\mu\nu}
+ \frac{1}{2} \bar{\lambda}^{a\alpha}(\gamma_{\mu})_{\alpha \beta}D^{ab}_{\mu}\lambda^{b\beta}
+ \frac{1}{2}\mathfrak{D}^{a}\mathfrak{D}^{a}
+ b^{a}\partial_{\mu}A^{a}_{\mu}
\nonumber \\
&&
+\bar{c}^{a}\Bigl[\partial_{\mu}D^{ab}_{\mu}c^{b}
-\bar{\epsilon}^{\alpha}(\gamma_{\mu})_{\alpha \beta}\partial_{\mu}\lambda^{a\beta}\Bigr]
+ T^{a}\Bigl[gf^{abc}c^{b}\mathfrak{D}^{c}
+ \bar{\epsilon}^{\alpha}(\gamma_{\mu})_{\alpha \beta}(\gamma_{5})^{\beta \eta} D_{\mu}^{ab}\lambda^{b}_{\eta}\Bigr]
\nonumber \\
&&
+L^{a}\Bigl[ \frac{g}{2}f^{abc}c^{b}c^{c}
-\bar{\epsilon}^\alpha(\gamma_{\mu})_{\alpha\beta}\epsilon^\beta A^{a}_{\mu}\Bigr]
- K^{a}_{\mu}\Bigl[D^{ab}_{\mu}c^{b}
- \bar{\epsilon}^\alpha(\gamma_\mu)_{\alpha\beta}\lambda^{a\beta}\Bigr]
- \Omega^{a}_{\mu} A^{a}_{\mu}
\nonumber\\
&&
+Y^{a\alpha}\Bigl[ gf^{abc}c^{b}\lambda^{c}_{\alpha} - \frac{1}{2}(\sigma_{\mu\nu})_{\alpha\beta} F_{\mu\nu}^{a}\epsilon^{\beta}
+ (\gamma_{5})_{\alpha\beta}\epsilon^{\beta} \mathfrak{D}^a\Bigr]
+ \Lambda^{a}c^{a}
- J^{a}\mathfrak{D}^{a}
+ X^{a\alpha}\lambda^{a}_{\alpha}
\biggl\}\,.
\end{eqnarray}
It remains now to introduce the  composite operators $A_{\mu}A_{\mu}$, $A_\mu \gamma_\mu \lambda$ and $\bar{\lambda}^{\alpha}\lambda_{\alpha}$ in a $Q$-invariant way. To that purpose, we shall make use of a second  set of external sources, $(j,\chi,\rho^{\alpha}_{\mu},\tau^{\alpha}_{\mu}, N, R)$, also assembled in $Q$-doublets, {\it i.e.}  
\begin{equation}
\left\{\begin{matrix}Q \chi = j\phantom{\Bigl|}\cr
Qj = \nabla \chi \phantom{\Bigl|} \end{matrix}\right.\,,\qquad
\left\{\begin{matrix}Q \tau^{\alpha}_{\mu} = \rho^{\alpha}_{\mu} \phantom{\Bigl|}\cr
Q \rho^{\alpha}_{\mu} = \nabla \tau^{\alpha}_{\mu} \phantom{\Bigl|}\end{matrix}\right.\,,\qquad
\left\{\begin{matrix}QR = N \phantom{\Bigl|}\cr
QN = \nabla R \phantom{\Bigl|}\end{matrix}\right.\,,
\end{equation}
and add to eq.\eqref{fullact} the following $Q$-invariant term:
\begin{eqnarray}
\Sigma_{AA-\bar{\lambda}\lambda} &=& Q\int d^{4}x\Bigg[\phantom{\Bigl|}\frac{1}{2}\,\chi\, A^{a}_{\mu}A^{a}_{\mu}+\frac{1}{2}\xi\chi j+ \tau^{\alpha}_{\mu}A_{\mu}^{a}\lambda^{a}_{\alpha}+R\bar{\lambda}^{a\alpha}\lambda_{\alpha}^{a} +\frac{\zeta}{4} R N^3 
\Bigg]\nonumber\\
&=&\int d^{4}x\,\Bigg[\frac{1}{2}jA_{\mu}^{a}A_{\mu}^{a} - \chi A_{\mu}^{a}\partial_{\mu}c^{a} + \chi A_{\mu}^{a}\bar{\epsilon}^{\alpha}[\gamma_{\mu}]_{\alpha\beta}\lambda^{a\beta} + \frac{\xi}{2}j^{2} - \frac{\xi}{2}\chi\bar{\epsilon}^{\alpha}[\gamma_{\mu}]_{\alpha\beta}\epsilon^{\beta}\partial_{\mu}\chi\nonumber\\
&&+\rho_{\alpha\mu}A^{a}_{\mu}\lambda^{a\alpha}-\tau_{\alpha\mu}(D_{\mu}^{ab}c^{b})\lambda^{a\alpha} + \tau_{\alpha\mu}\bar{\epsilon}^{\gamma}[\gamma_{\mu}]_{\gamma\beta}\lambda^{a\beta}\lambda^{a\alpha} + gf^{abc}\tau_{\alpha\mu}A_{\mu}^{a}c^{b}\lambda^{c\alpha}\nonumber\\
&&- \frac{1}{2}\tau_{\alpha\mu}A_{\mu}^{a}(\sigma_{\rho\nu})^{\alpha\beta}\epsilon_{\beta}F^{a}_{\rho\nu} +\tau_{\alpha\mu}A_{\mu}^{a}[\gamma_{5}]^{\alpha\beta}\epsilon_{\beta}\mathfrak D^{a}+N\left(\bar{\lambda}^{a\alpha}\lambda_{\alpha}^{a}\right) + \frac{\zeta}{4} N^4  \nonumber\\
&&+gf^{abc}Rc^{b}\bar{\lambda}^{c\alpha}\lambda_{\alpha}^{a}-\frac{R}{2}(\sigma_{\mu\nu})^{\alpha\gamma}\bar{\epsilon}_{\gamma}F_{\mu\nu}^{a}\lambda_{\alpha}^{a} +R(\gamma_{5})^{\alpha\gamma}\bar{\epsilon}_{\gamma}\mathfrak{D}^{a}\lambda_{\alpha}^{a}-gf^{abc}R\bar{\lambda}^{a\alpha}c^{b}\lambda_{\alpha}^{c}\nonumber \\
 &  &+\frac{R}{2}\bar{\lambda}^{a\alpha}(\sigma_{\mu\nu})_{\alpha\beta}\epsilon^{\beta}F_{\mu\nu}^{a}-R\bar{\lambda}^{a\alpha}(\gamma_{5})_{\alpha\beta}\epsilon^{\beta}\mathfrak{D}^{a}\Bigg]\,. \label{cop}
\end{eqnarray}
We see that both operators $A_{\mu}A_{\mu}$ and $\bar{\lambda}^{\alpha}\lambda_{\alpha}$ have been introduced through the corresponding external sources: the source $j$ for the operator $A_{\mu}A_{\mu}$ and the source $N$ for $\bar{\lambda}^{\alpha}\lambda_{\alpha}$. As one can figure out, the remaining terms in expression \eqref{cop}  are needed to keep full invariance under the operator $Q$. We also underline that the term quadratic in the external source $j$, {\it i.e.} $\frac{\xi}{2}j^{2}$, is allowed by power-counting, due to the fact both the external source $j$ and the operator $A_{\mu}A_{\mu}$ are of dimension two. As a consequence, the term $\frac{\xi}{2}j^{2}$ is of dimension four. Moreover, according to the local composite operator (LCO) procedure \cite{Verschelde:2001ia,Dudal:2002pq,Dudal:2003by,Dudal:2004rx}, this term is needed to take into account the ultraviolet divergences appearing in the correlator $\langle A^{2}(x)A^{2}(y)\rangle$, which will be  reabsorbed through the free dimensionless vacuum parameter $\xi$, as done in the non-supersymmetric case \cite{Verschelde:2001ia,Dudal:2002pq,Dudal:2003by,Dudal:2004rx}.  Analogously, the vacuum term $\frac{\zeta}{4} N^4$ plays the same role, {\it i.e.} it is needed to account for the ultraviolet divergences which might arise in the correlation functions of the composite operator $\bar{\lambda}^{\alpha}\lambda_{\alpha}$, which will be reabsorbed through the free dimensionless vacuum parameter ${\zeta}$. Notice also that, due to the fact that the operator $\bar{\lambda}^{\alpha}\lambda_{\alpha}$ has dimension three, the corresponding source $N$ has dimension one, explaining therefore the presence of the dimension four term $\frac{\zeta}{4} N^4$. Finally, let us observe that the external source $\chi$ takes into account the presence of the composite operator $A_\mu \gamma_\mu \lambda$. \\\\Therefore, for the classical  complete classical action $\Sigma$  including all above-mentioned composite operators, we have
\begin{eqnarray}
\Sigma & = & \int d^{4}x\Bigg[\frac{1}{4}F_{\mu\nu}^{a}F_{\mu\nu}^{a}+\frac{1}{2}\bar{\lambda}^{a\alpha}[\gamma_{\mu}]_{\alpha\beta}D_{\mu}^{ab}\lambda^{b\beta}+\frac{1}{2}\mathfrak{D^{a}}\mathfrak{D^{a}}+\bar{c}^{a}\partial_{\mu}D_{\mu}^{ab}c^{b}+b^{a}\partial_{\mu}A_{\mu}^{a}\nonumber \\
 &  & -\bar{c}^{a}\bar{\epsilon}^{\alpha}[\gamma_{\mu}]_{\alpha\beta}\partial_{\mu}\lambda^{a\beta}+\Lambda^{a}c^{a}+L^{a}\big[\frac{1}{2}gf^{abc}c^{b}c^{c}-\bar{\epsilon}^{\alpha}[\gamma_{\mu}]_{\alpha\beta}\epsilon^{\beta}A_{\mu}^{a}\big]-J^{a}\mathfrak{D}^{a}\nonumber \\
 &  & -\Omega_{\mu}^{a}A_{\mu}^{a}+K_{\mu}^{a}\Big[-D_{\mu}^{ab}c^{b}+\bar{\epsilon}^{\alpha}[\gamma_{\mu}]_{\alpha\beta}\lambda^{a\beta}\Big]\nonumber \\
 &  & +T^{a}\big[gf^{abc}c^{b}\mathfrak{D}^{c}-\bar{\epsilon}^{\alpha}[\gamma_{\mu}]_{\alpha\beta}D_{\mu}^{ab}[\gamma_{5}]^{\beta\eta}\lambda_{\eta}^{b}\big]+X^{a\alpha}\lambda_{\alpha}^{a}+gf^{abc}Y^{a\alpha}c^{b}\lambda_{\alpha}^{c}\nonumber \\
 &  & +Y^{a\alpha}\big[-\frac{1}{2}(\sigma_{\mu\nu})_{\alpha\beta}\epsilon^{\beta}F_{\mu\nu}^{a}+[\gamma_{5}]_{\alpha\beta}\epsilon^{\beta}\mathfrak{D}^{a}\big]+\rho_{\alpha\mu}A_{\mu}^{a}\lambda^{a\alpha}\nonumber \\
 &  & -\tau_{\alpha\mu}(D_{\mu}^{ab}c^{b})\lambda^{a\alpha}+\tau_{\alpha\mu}\bar{\epsilon}^{\gamma}[\gamma_{\mu}]_{\gamma\beta}\lambda^{a\beta}\lambda^{a\alpha}+gf^{abc}\tau_{\alpha\mu}A_{\mu}^{a}c^{b}\lambda^{c\alpha}-\frac{1}{2}\tau_{\alpha\mu}A_{\mu}^{a}(\sigma_{\rho\nu})^{\alpha\beta}\epsilon_{\beta}F_{\rho\nu}^{a}+\nonumber \\
 &  & +\tau_{\alpha\mu}A_{\mu}^{a}[\gamma_{5}]^{\alpha\beta}\epsilon_{\beta}\mathfrak{D}^{a}+\frac{1}{2}jA_{\mu}^{a}A_{\mu}^{a}-\chi A_{\mu}^{a}\partial_{\mu}c^{a}+\chi A_{\mu}^{a}\bar{\epsilon}^{\alpha}[\gamma_{\mu}]_{\alpha\beta}\lambda^{a\beta}+\nonumber \\
 &  & +\frac{\xi}{2}j^{2}-\frac{\xi}{2}\chi\bar{\epsilon}^{\alpha}[\gamma_{\mu}]_{\alpha\beta}\epsilon^{\beta}\partial_{\mu}\chi+N\left(\bar{\lambda}^{a\alpha}\lambda_{\alpha}^{a}\right)+gf^{abc}Rc^{b}\bar{\lambda}^{c\alpha}\lambda_{\alpha}^{a}-\frac{R}{2}(\sigma_{\mu\nu})^{\alpha\gamma}\bar{\epsilon}_{\gamma}F_{\mu\nu}^{a}\lambda_{\alpha}^{a}\nonumber \\
 &  & +R(\gamma_{5})^{\alpha\gamma}\bar{\epsilon}_{\gamma}\mathfrak{D}^{a}\lambda_{\alpha}^{a}-gf^{abc}R\bar{\lambda}^{a\alpha}c^{b}\lambda_{\alpha}^{c}+\frac{R}{2}\bar{\lambda}^{a\alpha}(\sigma_{\mu\nu})_{\alpha\beta}\epsilon^{\beta}F_{\mu\nu}^{a} + \frac{\zeta}{4} N^4 \nonumber \\
 &  & -R\bar{\lambda}^{a\alpha}(\gamma_{5})_{\alpha\beta}\epsilon^{\beta}\mathfrak{D}^{a}\Bigg].
\label{SSYM}
\end{eqnarray}
This action will be taken as the starting point  for the algebraic renormalization analysis \cite{Piguet:1995er}.  

\section{Ward identities and algebraic characterization of the invariant counterterm}

Let us start by first establishing all Ward identities fulfilled by the complete classical action $\Sigma$, eq.\eqref{SSYM}, a task which we shall face in the next sub-section. 

\subsection{Ward identities}

It turns out that the complete action $\Sigma$ obeys a large set of Ward identities, enlisted below, namley: 

\begin{itemize}
{\item The Slavnov-Taylor identity:}

\begin{equation}
\mathcal{S}(\Sigma) = 0 \;, \label{STid}
\end{equation}
where
\begin{eqnarray}
\mathcal{S}\Big(\Sigma\Big) &=& \int d^{4}x \Bigg[\frac{\delta \Sigma}{\delta K^{a}_{\mu}}\frac{\delta \Sigma}{\delta A^{a}_{\mu}} + \Omega_{\mu}^{a}\frac{\delta \Sigma}{\delta K^{a}_{\mu}} + \frac{\delta \Sigma}{\delta \lambda^{a\alpha}}\frac{\delta \Sigma}{\delta Y^{a}_{\alpha}} + X^{a\alpha}\frac{\delta \Sigma}{\delta Y^{a\alpha}} + \frac{\delta \Sigma}{\delta c^{a}}\frac{\delta \Sigma}{\delta L^{a}} + \Lambda^{a}\frac{\delta \Sigma}{\delta L^{a}}\nonumber\\
&+& \frac{\delta \Sigma}{\delta \mathfrak D^{a}}\frac{\delta \Sigma}{\delta T^{a}} + J^{a}\frac{\delta \Sigma}{\delta T^{a}} - X^{a\alpha}[\gamma_{5}]_{\alpha\beta}\epsilon^{\beta}\frac{\delta \Sigma}{\delta \mathfrak D^{a}} + b^{a}\frac{\delta \Sigma}{\delta \bar{c}^{a}} + \Big(\nabla \bar{c}^{a}\Big)\frac{\delta \Sigma}{\delta b^{a}} + \Big(\nabla K^{a}_{\mu}\Big)\frac{\delta \Sigma}{\delta \Omega^{a}_{\mu}}\nonumber\\
&+& \Big(\nabla Y^{a\alpha}\Big)\frac{\delta \Sigma}{\delta X^{a\alpha}} + \Big(\nabla T^{a}\Big)\frac{\delta \Sigma}{\delta J^{a}} + \Big(\nabla L^{a}\Big)\frac{\delta \Sigma}{\delta \Lambda^{a}} + j\frac{\delta \Sigma}{\delta \chi} + \Big(\nabla \chi\Big)\frac{\delta \Sigma}{\delta j} +\nonumber\\
&+& \rho^{\alpha}_{\mu}\frac{\delta \Sigma}{\delta \tau^{\alpha}_{\mu}} + \Big(\nabla \tau^{\alpha}_{\mu}\Big)\frac{\delta \Sigma}{\delta \rho^{\alpha}_{\mu}}+N\frac{\delta\Sigma}{\delta R}+\Big(\nabla R\Big)\frac{\delta\Sigma}{\delta N}\Bigg]\,.
\label{ST}
\end{eqnarray}

Let us also introduce, for later convenience, the so-called linearized Slavnov-Taylor operator $\mathcal{B}_{\Sigma}$ \cite{Piguet:1995er}, defined as:
\begin{eqnarray}
\mathcal{B}_{\Sigma} &=& \int d^{4}x \Bigg[\frac{\delta \Sigma}{\delta K^{a}_{\mu}}\frac{\delta }{\delta A^{a}_{\mu}} + \frac{\delta \Sigma}{\delta A^{a}_{\mu}}\frac{\delta }{\delta K^{a}_{\mu}}  +\Omega_{\mu}^{a}\frac{\delta}{\delta K^{a}_{\mu}} + \frac{\delta \Sigma}{\delta \lambda^{a\alpha}}\frac{\delta}{\delta Y^{a}_{\alpha}} + X^{a\alpha}\frac{\delta}{\delta Y^{a}_{\alpha}} +\nonumber\\
&+& \frac{\delta \Sigma}{\delta Y^{a\alpha}}\frac{\delta }{\delta \lambda^{a}_{\alpha}} + \frac{\delta \Sigma}{\delta c^{a}}\frac{\delta}{\delta L^{a}} +\frac{\delta \Sigma}{\delta L^{a}}\frac{\delta }{\delta c^{a}} + \Lambda^{a}\frac{\delta}{\delta L^{a}} + \frac{\delta \Sigma}{\delta \mathfrak D^{a}}\frac{\delta}{\delta T^{a}} + \frac{\delta \Sigma}{\delta T^{a}}\frac{\delta}{\delta \mathfrak D^{a}}+\nonumber\\
&+& J^{a}\frac{\delta}{\delta T^{a}} - X^{a\alpha}[\gamma_{5}]_{\alpha\beta}\epsilon^{\beta}\frac{\delta}{\delta \mathfrak D^{a}} + J^{a}[\gamma_{5}]_{\alpha\beta}\epsilon^{\beta}\frac{\delta}{\delta \lambda^{a\alpha}} + b^{a}\frac{\delta}{\delta \bar{c}^{a}} + \Big(\nabla \bar{c}^{a}\Big)\frac{\delta}{\delta b^{a}}+\nonumber\\ 
&+&\Big(\nabla K^{a}_{\mu}\Big)\frac{\delta}{\delta \Omega^{a}_{\mu}} + \Big(\nabla Y^{a\alpha}\Big)\frac{\delta}{\delta X^{a\alpha}} + \Big(\nabla T^{a}\Big)\frac{\delta}{\delta J^{a}} + \Big(\nabla L^{a}\Big)\frac{\delta}{\delta \Lambda^{a}} + j\frac{\delta}{\delta \chi} +\nonumber\\
&+&\Big(\nabla \chi\Big)\frac{\delta}{\delta j} + \rho^{\alpha}_{\mu}\frac{\delta}{\delta \tau^{\alpha}_{\mu}} + \Big(\nabla \tau^{\alpha}_{\mu}\Big)\frac{\delta}{\delta \rho^{\alpha}_{\mu}}+N\frac{\delta}{\delta R}+\Big(\nabla R\Big)\frac{\delta}{\delta N}\Bigg]\,,
\label{LST}
\end{eqnarray}
which has the following property 
\begin{equation}
\mathcal{B}_{\Sigma}  \mathcal{B}_{\Sigma}  = \nabla \;. \label{bnilp}
\end{equation}
As a consequence, $\mathcal{B}_{\Sigma}$ is nilpotent when acting on space-time integrated local functionals of the fields, sources and their space-time derivatives.  
\item{The Landau gauge-fixing condition and the  anti-ghost equation \cite{Piguet:1995er}:}
\begin{equation}
\frac{\delta\Sigma}{\delta b^{a}}= i\partial_{\mu}A^{a}_{\mu}\,,\qquad
\frac{\delta\Sigma}{\delta\bar{c}^{a}}+\partial_{\mu}\frac{\delta\Sigma}{\delta K^{a}_{\mu}}=0\,.
\label{GFandAntiGhost}
\end{equation}

\item{The Landau ghost Ward identity  \cite{Piguet:1995er,Blasi:1990xz}:}
\begin{equation}
G^{a}(\Sigma)=\Delta^{a}_{\mathrm{class}}\,,  \label{gW}
\end{equation}
where
\begin{equation}
G^{a}=\int d^{4}x\,\biggl[\frac{\delta}{\delta{c}^{a}} 
+ gf^{abc}\bar{c}^{b}\frac{\delta}{\delta{b}^{c}}\biggl]\,,
\end{equation}
and
\begin{equation}
\Delta^{a}_{\mathrm{class}}=\int d^{4}x\,\left[gf^{abc}\left(K^{b}_{\mu}A^{c}_{\mu}
-L^{b}c^{c}+T^{b}\mathfrak{D}^{c}
-Y^{b\alpha}\lambda^{c}_{\alpha}\right)-\Lambda^{a}\right]\,.    \label{bex}
\end{equation}
It's important to notice that the term $\Delta^{a}_{\mathrm{class}}$, eqs.\eqref{gW},\eqref{bex}, is purely linear in the quantum fields. As such, $\Delta^{a}_{\mathrm{class}}$ is a classical breaking, not affected by the quantum corrections \cite{Piguet:1995er,Blasi:1990xz}.

\item{The equation of motion of the auxiliary field $\mathfrak{D}^{a}$:}

\begin{eqnarray}
\frac{\delta\Sigma}{\delta{\mathfrak{D}^{a}}}&=&  \mathfrak{D}^{a} 
- J^{a} + gf^{abc}c^{b}T^{c} + Y^{a\alpha}(\gamma_{5})_{\alpha\beta}\,\varepsilon^{\beta}- \tau^{\alpha}_{\mu}A_{\mu}^{a}(\gamma_{5})_{\alpha\beta}\epsilon^{\beta}\nonumber\\
&&+R(\gamma_{5})^{\alpha\gamma}\bar{\epsilon}_{\gamma}\lambda_{\alpha}^{a}+R\bar{\lambda}^{a\alpha}(\gamma_{5})_{\alpha\beta}\epsilon^{\beta}\,.  \label{auxW}
\end{eqnarray}
Notice that the right-hand side of equation \eqref{auxW} is also linear in the quantum fields, being again a classical breaking.  

\item{The linearly broken Ward identity of the gluino:}

\begin{equation}
\label{eqT}
\frac{\delta\Sigma }{\delta T^{a}} - gf^{abc}\left(c^{b}\frac{\delta\Sigma}{\delta \mathfrak D^{c}}-T^{b}\frac{\delta\Sigma}{\delta L^{c}}\right) + [\gamma_{5}]^{\alpha}_{\beta}\epsilon^{\beta}\left(\frac{\delta\Sigma}{\delta \lambda^{a\alpha}} + \tau_{\alpha\mu}\frac{\delta\Sigma }{\delta K^{a}_{\mu}}+R\frac{\delta\Sigma}{\delta Y_{\alpha}^{a}}\right)=\tilde\Delta^{a}_{\mathrm{class}}\,,
\end{equation}
where  $\tilde\Delta^{a}_{\mathrm{class}}$ is a classical breaking, namely 
\begin{eqnarray}
\tilde{\Delta}^{a}_{\mathrm{class}} &=&
(\gamma_{5})^{\alpha\beta}\epsilon_{\beta}\Big[\partial_{\mu}\bar{c}^{a} + K^{a}_{\mu} + \chi A^{a}_{\mu}\Big]\bar{\epsilon}^{\gamma}(\gamma_{\mu})_{\gamma\alpha} + (\gamma_{5})^{\alpha\beta}\epsilon_{\beta}\Big[\rho_{\alpha\mu}A^{a}_{\mu} + X^{a}_{\alpha}\Big]  \nonumber\\ 
&&+ 2gf^{abc}T^{b}A^{c}_{\mu}\Big[\bar{\epsilon}^{\alpha}(\gamma_{\mu})_{\alpha\beta}\epsilon^{\beta}\Big] + \nabla T^{a} - gf^{abc}c^{b}J^{c}
-N(\gamma_{5})_{\alpha\beta}\epsilon^{\beta}\bar{\lambda}^{a\alpha}\,.
\end{eqnarray}

In particular, it can be noticed that the Ward identity of the gluino, eq.\eqref{eqT}, can be  originated through the commutation  between the Slavnov-Taylor identity, eq.\eqref{ST}, and the Ward identity  \eqref{auxW}. 

\item{The equations of motion of the external sources $J^{a}$ and $\Lambda^{a}$:}
\begin{equation}
\frac{\delta \Sigma}{\delta \Lambda^{a}} = c^{a}\,,\qquad
\frac{\delta \Sigma}{\delta J^{a}} = -\mathfrak D^{a}\,.
\end{equation}

\item{The Ward identity for the sources $\chi$ and $\rho$:}
\begin{eqnarray}
\int d^{4}x \Bigg[\frac{\delta \Sigma}{\delta \chi} + c^{a}\frac{\delta \Sigma}{\delta b^{a}} - \bar{\epsilon}^{\beta}(\gamma_{\mu})_{\beta\alpha}\frac{\delta \Sigma}{\delta \rho_{\alpha\mu}}\Bigg] = 0\,. 
\end{eqnarray}

\end{itemize}

\subsection{Discrete Symmetries}
\hspace{0.5 cm}Besides the Ward identities of the previous subsection, the action \eqref{SSYM} is left invariant by two useful discrete symmetries. Firstly, let us change $x_{4} \to -x_{4}$ and
\begin{equation}
\gamma_{4} \rightarrow -\gamma_{4}\,,\qquad \gamma_{k} \rightarrow \gamma_{k}\,,\qquad k = 1,2,3\,.
\label{44}
\end{equation}
Notice that the anti-commutation relation $\lbrace\gamma_{\mu},\gamma_{\nu}\rbrace=2\delta_{\mu\nu}$ remains unchanged by the transformations above, while
\begin{equation}
\gamma_{5} \rightarrow -\gamma_{5}\,,\qquad
\mathcal{C}\rightarrow -\mathcal{C}\,,\qquad
\sigma_{4k} \rightarrow -\sigma_{4k}\,,\qquad 
\sigma_{kl} \rightarrow \sigma_{kl}\,,\qquad
k, l = 1,2,3\,.
\label{45}
\end{equation}
As a consequence, the action $\Sigma$ turns out to be left invariant by the following discrete transformations of fields and sources:
\begin{eqnarray}
(A^{a}_{4},\mathfrak D^{a}, K^{a}_{4}, \Omega^{a}_{4}, J^{a}, T^{a} )
 &\longrightarrow& -(A^{a}_{4}, \mathfrak D^{a}, K^{a}_{4}, \Omega^{a}_{4}, J^{a}, T^{a}), \nonumber\\
(\lambda^{a\alpha}, \tau^{\alpha}_{4}, \epsilon^{\alpha}) &\longrightarrow& i(\lambda^{a\alpha}, \tau^{\alpha}_{4}, \epsilon^{\alpha} ), \nonumber\\
(\bar{\lambda}^{a\alpha}, X^{a\alpha}, Y^{a\alpha}, \rho^{\alpha}_{4}, \bar{\epsilon}^{\alpha}) &\longrightarrow& -i(\bar{\lambda}^{a\alpha}, X^{a\alpha}, Y^{a\alpha}, \rho^{\alpha}_{4}, \bar{\epsilon}^{\alpha})\,.
\label{46}
\end{eqnarray}
A second discrete symmetry is obtained by letting $x_{1} \rightarrow -x_{1}$ and
\begin{equation}
\gamma_{1} \rightarrow -\gamma_{1}\,,\qquad 
\gamma_{k} \rightarrow \gamma_{k}\,,\qquad
k = 2,3,4\,.
\label{47}
\end{equation}
Also here the anti-commutation relation between the $\gamma$ matrices remains unchanged, while
\begin{equation}
\gamma_{5} \rightarrow -\gamma_{5}\,,\qquad
\mathcal{C}\rightarrow \mathcal{C}\,,\qquad
\sigma_{1k} \rightarrow -\sigma_{1k}\,,\qquad 
\sigma_{kl} \rightarrow \sigma_{kl}\,,\qquad
k, l = 2,3,4\,.
\label{48}
\end{equation}
Again, the action $\Sigma$ turns out to be left invariant by the following set of discrete transformations:
\begin{eqnarray}
(A^{a}_{1},\mathfrak D^{a}, K^{a}_{1}, \Omega^{a}_{1}, J^{a}, T^{a},\tau^{\alpha}_{1},\rho^{\alpha}_{1})
 &\longrightarrow& -(A^{a}_{1},\mathfrak D^{a}, K^{a}_{1}, \Omega^{a}_{1}, J^{a}, T^{a},\tau^{\alpha}_{1},\rho^{\alpha}_{1})\,.
\label{49}
\end{eqnarray}

\section{The algebraic characterization of the invariant counterterm and renormalizability}

Having established the Ward identities fulfilled by the action $\Sigma$, we are now ready to characterize the most general local invariant counterterm which can be freely added to each order in perturbation theory.  Following the  algebraic renormalization procedure \cite{Piguet:1995er},  we perturb the complete action $\Sigma$ by adding an arbitrary  local integrated polynomial in the fields and sources, $\Sigma_{count}$, with dimension four and vanishing ghost number and we require that the perturbed action,  {\it i.e.} $(\Sigma + \omega \Sigma_{count})$, obeys the same Ward identities of the starting action $\Sigma$ to the first order in the infinitesimal expansion parameter $\omega$.  In other words, one imposes that 
\begin{equation} 
\mathcal{S}(\Sigma + \omega \Sigma_{count}) = 0 + O(\omega^2)\;, \label{stp}
\end{equation}
\begin{equation}
\frac{\delta(\Sigma+ \omega \Sigma_{count})}{\delta b^{a}}= \partial_{\mu}A^{a}_{\mu} + O(\omega^2) \,,\qquad
\left( \frac{\delta}{\delta\bar{c}^{a}}+\partial_{\mu}\frac{\delta} {\delta K^{a}_{\mu}}\right) (\Sigma+ \omega \Sigma_{count}) =0+ O(\omega^2) \,,
\label{GFandAntiGhostp}
\end{equation}
\begin{equation}
G^{a}(\Sigma+ \omega \Sigma_{count})=\Delta^{a}_{\mathrm{class}}+O(\omega^2)\,,  \label{gWp}
\end{equation}
\newpage
\begin{eqnarray}
\frac{\delta(\Sigma+ \omega \Sigma_{count})}{\delta{\mathfrak{D}^{a}}}&=&  \mathfrak{D}^{a} 
- J^{a} + gf^{abc}c^{b}T^{c} + Y^{a\alpha}(\gamma_{5})_{\alpha\beta}\,\varepsilon^{\beta}- \tau^{\alpha}_{\mu}A_{\mu}^{a}(\gamma_{5})_{\alpha\beta}\epsilon^{\beta}\nonumber\\
&&+R(\gamma_{5})^{\alpha\gamma}\bar{\epsilon}_{\gamma}\lambda_{\alpha}^{a}+R\bar{\lambda}^{a\alpha}(\gamma_{5})_{\alpha\beta}\epsilon^{\beta} +O(\omega^2)\,,  \label{auxWp}
\end{eqnarray}
\begin{eqnarray}
\label{eqTp}
\left[\frac{\delta }{\delta T^{a}} - gf^{abc}\left(c^{b}\frac{\delta}{\delta \mathfrak D^{c}}-T^{b}\frac{\delta}{\delta L^{c}}\right) + [\gamma_{5}]^{\alpha}_{\beta}\epsilon^{\beta}\left(\frac{\delta}{\delta \lambda^{a\alpha}} + \tau_{\alpha\mu}\frac{\delta }{\delta K^{a}_{\mu}}+R\frac{\delta}{\delta Y_{\alpha}^{a}}\right)\right](\Sigma + \omega \Sigma_{count}) &=& \nonumber\\ \tilde\Delta^{a}_{\mathrm{class}} + O(\omega^2) \;. 
\end{eqnarray}
\begin{equation}
\label{eqTp2}
\frac{\delta (\Sigma + \omega \Sigma_{count})}{\delta \Lambda^{a}} = c^{a}+ O(\omega^2)\,,\qquad
\frac{\delta (\Sigma + \omega \Sigma_{count})}{\delta J^{a}} = -\mathfrak D^{a}+ O(\omega^2)\,.
\end{equation}
\begin{eqnarray}
\int d^{4}x \Bigg[\frac{\delta }{\delta \chi} + c^{a}\frac{\delta }{\delta b^{a}} - \bar{\epsilon}^{\beta}(\gamma_{\mu})_{\beta\alpha}\frac{\delta }{\delta \rho_{\alpha\mu}}\Bigg](\Sigma + \omega \Sigma_{count}) = 0 + O(\omega^2)\,. 
\label{eqTp3}
\end{eqnarray}
To the  first order in the expansion parameter $\omega$, equations \eqref{stp}, \eqref{GFandAntiGhostp},  \eqref{gWp}, \eqref{auxWp}, \eqref{eqTp}, \eqref{eqTp2} and \eqref{eqTp3} give the  following conditions on  $\Sigma_{count}$:
\begin{equation}
\mathcal{B}_{\Sigma} (\Sigma_{count}) = 0\;,  \label{c1}
\end{equation}
\begin{equation}
\frac{\delta}{\delta b^{a}}\,\Sigma_{count}=0 \;, \qquad  \left( \frac{\delta}{\delta\bar{c}^{a}} + \partial_\mu\frac{\delta}{\delta K_\mu^{a}} \right) \Sigma_{count}=0\;, \label{c2}
\end{equation}
\begin{equation}
G^{a}\,\Sigma_{count}=0  \;, \label{c3}
\end{equation}
\begin{equation}
\frac{\delta}{\delta \mathfrak{D}^{a}}\,\Sigma_{count}=0  \;, \label{c4}
\end{equation} 
\begin{equation}
\label{c5}
\left[\frac{\delta }{\delta T^{a}} - gf^{abc}\left(c^{b}\frac{\delta}{\delta \mathfrak D^{c}}-T^{b}\frac{\delta}{\delta L^{c}}\right) + [\gamma_{5}]^{\alpha}_{\beta}\epsilon^{\beta}\left(\frac{\delta}{\delta \lambda^{a\alpha}} + \tau_{\alpha\mu}\frac{\delta }{\delta K^{a}_{\mu}}+R\frac{\delta}{\delta Y_{\alpha}^{a}}\right)\right]\Sigma_{count} =  0 \;, 
\end{equation}
\begin{equation}
\label{eqc9}
\frac{\delta \Sigma_{count}}{\delta \Lambda^{a}} = 0 \,,\qquad
\frac{\delta \Sigma_{count}}{\delta J^{a}} = 0 \,.
\end{equation}
\begin{eqnarray}
\int d^{4}x \Bigg[\frac{\delta }{\delta \chi} + c^{a}\frac{\delta }{\delta b^{a}} - \bar{\epsilon}^{\beta}(\gamma_{\mu})_{\beta\alpha}\frac{\delta }{\delta \rho_{\alpha\mu}}\Bigg]\Sigma_{count} = 0,
\label{c10} 
\end{eqnarray}
where $\mathcal{B}_{\Sigma}$ is the linearized Salvonv-Taylor operator of eq.\eqref{LST}. \\\\In particular, the first condition, eq.\eqref{c1}, tell us that the counterterm  $\Sigma_{count}$ belongs to the cohomolgy of the operator $\mathcal{B}_{\Sigma}$  in the space of the local integrated polynomials in the fields and external sources with dimension four. From the general results on the  cohomology of Yang-Mills theories, see \cite{Piguet:1995er}, it follows that $\Sigma_{count}$ can be parametrized in the following way
\begin{equation}
\label{cttnt}
\Sigma_{count}=a_{0}\,S_{\mathrm{SYM}} + \mathcal{B}_{\Sigma} \Delta^{(-1)} \;,
\end{equation}
where $a_0$ is a free coefficient and $\Delta^{(-1)}$ is the most general local integrated polynomial in the fields and external sources with ghost number $-1$ and dimension $3$. \\\\From the tables displayed in Appendix A, the most general expression for $\Delta^{(-1)}$ is given by:
\begin{eqnarray}
\Delta^{(-1)} & = & \int d^{4}x\Bigg[\Big(a_{1}\partial_{\mu}\bar{c}^{a}+a_{2}K_{\mu}^{a}\Big)A_{\mu}^{a}+a_{3}\chi A_{\mu}^{a}A_{\mu}^{a}+a_{4}\xi j\chi+a_{5}\tau_{\alpha\mu}A_{\mu}^{a}\lambda^{a\alpha}\nonumber\\
&&+ a_{6}[\gamma_{5}]_{\alpha\beta}\tau_{\mu}^{\alpha}\epsilon^{\beta}A_{\mu}^{a}T^{a}+a_{7}\mathfrak{D}^{a}
\bar{c}^{a}+a_{8}\mathfrak{D}^{a}T^{a}+a_{9}\bar{\lambda}^{a\alpha}\tau_{\mu\alpha}A_{\mu}^{a}+\nonumber \\ &&+ a_{10}J^{a}T^{a}+a_{11}Y_{\alpha}^{a}\lambda^{a\alpha}+a_{12}Y^{a\alpha}[\gamma_{5}]_{\alpha\beta}\epsilon^{\beta}T^{a}+a_{13}\bar{\lambda}^{a\alpha}Y_{\alpha}^{a}
\nonumber \\&&+a_{14}\bar{\epsilon}^{\beta}[\gamma_{5}]_{\beta\alpha}T^{a}Y^{a\alpha}+a_{15}L^{a}c^{a}+a_{16}gf^{abc}T^{a}T^{b}\bar{c}^{c}
+a_{17}\bar{c}^{a}\bar{\epsilon}^{\alpha}[\gamma_{5}]_{\alpha\beta}Y^{a\beta}\nonumber \\
 &&+a_{18}gf^{abc}c^{a}\bar{c}^{b}T^{c}+a_{19}\tau_{\mu}^{\alpha}[\sigma_{\mu\nu}]_{\alpha\beta}A_{\nu}^{a}\lambda^{a\beta}+a_{20}\bar{\lambda}^{a\alpha}[\sigma_{\mu\nu}]_{\alpha\beta}\tau_{\mu}^{\beta}A_{\nu}^{a}\nonumber\\
&&+ a_{21}Y^{a\alpha}[\gamma_{5}]_{\alpha\beta}\lambda^{a\beta}+a_{22}\bar{\lambda}^{a\alpha}[\gamma_{5}]_{\alpha\beta}Y^{a\beta}+a_{23}R\bar{\lambda}^{a\alpha}\lambda^{a}_{\alpha}
+\frac{a_{24}}{4}RN^{3}\Bigg]
 \,,
\end{eqnarray}
with $a_{i}$, $(i= 1,...,24)$,  arbitrary coefficients. It's worth to mention that, according to Appendix A, we have chosen dimension $1$  for both Fadeev-Popov ghost and anti-ghost fields. This convenient choice simplifies the analysis of the counterterm, as enables us  to assign dimension ($\frac{1}{2}$) to the susy global ghost parameter $\epsilon^{\alpha}$. \\\\Applying now the discrete symmetries, eqs.\eqref{44}, \eqref{45}, \eqref{46}, \eqref{47}, \eqref{48}, \eqref{49}, it turns out that 
\begin{equation}
a_{7} = a_{9} = a_{13} = a_{14} = a_{18} = a_{19} = a_{20} = a_{21} = 0.
\end{equation}
From now on, we will make use of the constraints \eqref{c1} -- \eqref{c10} in combination with \eqref{cttnt} in order to determining  the remaining independent coefficients. Though, before this, it might be useful to establish some interesting commuting and anti-commuting relations among the various functional operators corresponding to the previous Ward identities, namely: 
\begin{eqnarray}
&&\Bigg\{\frac{\delta}{\delta\bar{c}^{a}} + \partial_{\mu}\frac{\delta}{\delta K_{\mu}^{a}}\,\,,\,\,\mathcal{B}_{\Sigma}\Bigg\} \,\,=\,\, -\nabla\left(\frac{\delta}{\delta b^{a}} + \partial_{\mu}\frac{\delta}{\delta \Omega_{\mu}^{a}}\right)\,,\\
&&\left\{\int d^{4}x\frac{\delta}{\delta\chi} + c^{a}\frac{\delta}{\delta b^{a}} - \bar{\epsilon}^{\beta}[\gamma_{\mu}]_{\beta\alpha}\frac{\delta}{\delta\rho_{\alpha\mu}}\,\, ,\,\,\mathcal{B}_{\Sigma}\right\}\,\, =\,\, \int d^{4}x \Bigg[c^{a}\frac{\delta}{\delta \bar{c}^{a}}-\nabla\frac{\delta}{\delta j} - \frac{\delta\Sigma}{\delta b^{a}}\frac{\delta}{\delta L^{a}}\nonumber\\
&&\phantom{\left\{\int d^{4}x\frac{\delta}{\delta\chi} + c^{a}\frac{\delta}{\delta b^{a}} - \bar{\epsilon}^{\beta}[\gamma_{\mu}]_{\beta\alpha}\frac{\delta}{\delta\rho_{\alpha\mu}}\,\, ,\,\,\mathcal{B}_{\Sigma}\right\}\,\, =\,\, }+\frac{\delta \Sigma}{\delta L^{a}}\frac{\delta}{\delta b^{a}} - \bar{\epsilon}^{\beta}[\gamma_{\mu}]_{\beta\alpha}\frac{\delta}{\delta\tau_{\alpha\mu}}\Bigg] \,, \label{cc1}
\end{eqnarray}
\begin{eqnarray}
\Bigg[\frac{\delta}{\delta \mathfrak D^{a}}\,\,,\,\,\mathcal{B}_{\Sigma}\Bigg] \,\,=\,\, \frac{\delta }{\delta T^{a}} - gf^{abc}\left(c^{b}\frac{\delta}{\delta \mathfrak D^{c}}-T^{b}\frac{\delta}{\delta L^{c}}\right) + [\gamma_{5}]^{\alpha}_{\beta}\epsilon^{\beta}\left(\frac{\delta}{\delta \lambda^{a\alpha}} + \tau_{\alpha\mu}\frac{\delta }{\delta K^{a}_{\mu}}+R\frac{\delta}{\delta Y_{\alpha}^{a}}\right)\,. \label{cc2} 
\end{eqnarray}
Applying thus the equations \eqref{cc1} and \eqref{cc2} to the counterterm $\Sigma_{count}$, the following  result is found for $\Delta^{(-1)}$:
\begin{eqnarray}
\Delta^{(-1)} &=& \int d^{4}x \Bigg[a_{1}\Big(\partial_{\mu}\bar{c}^{a} + K^{a}_{\mu}\Big)A^{a}_{\mu}  -\frac{1}{2}a_{1}\Big(\chi A^{a}_{\mu}A^{a}_{\mu} + 2\tau_{\alpha\mu}A^{a}_{\mu}\lambda^{a\alpha}\Big)+\nonumber \\
&+& a_{4}\xi j\chi - \frac{a_{0}}{2}\mathfrak D^{a}T^{a} + a_{11}Y^{a}_{\alpha}\lambda^{a\alpha} + \Big(\frac{a_{0}}{2} - a_{11}\Big)Y^{a\alpha}[\gamma_{5}]_{\alpha\beta}\epsilon^{\beta}T^{a}\nonumber\\
&+&a_{23}R\bar{\lambda}^{a\alpha}\lambda^{a}_{\alpha}
+\frac{a_{24}}{4}RN^{3}\Bigg]\,,\label{fct} 
\end{eqnarray}
where,
\begin{eqnarray}
a_{6} &=& a_{10} = a_{15} = a_{16} = a_{17} = a_{22}  = 0, \nonumber\\ 
a_{1} &=& a_{2},\nonumber\\
a_{3} &=& -\frac{1}{2}a_{1},\nonumber\\
a_{5} &=& 2a_{3},\nonumber\\
a_{8} &=& \frac{-a_{0}}{2},\nonumber\\ 
a_{12} &=& \frac{a_{0}}{2} - a_{11}.
\end{eqnarray}
We observe that $\Sigma_{count}$ contains six arbitrary coefficients $a_{0}$, $a_{1}$, $a_{4}$, $a_{11}$, $a_{23}$  and $a_{24}$, which will give rise to the renormalization factors of all fields, sources and parameters. To complete the algebraic renormalization analysis of the model, we need to show  that the counterterm $\Sigma_{count}$ can be re-absorbed in the starting action through a redefinition of the fields and parameters $\{\varphi \}$, $\varphi = (A,\lambda, b, c, \bar {c}, \mathfrak{D}, \epsilon)$, of the external sources $\{ \Phi \}$, $\Phi= (K,\Omega, \Lambda, T, J, L, Y, X, j,\rho,\tau, N, R)$, of the coupling constant $g$ and of the vacuum parameters $({\xi}, \zeta)$, namely:
\begin{equation}
\label{ration}
\Sigma(\varphi,\Phi,g,\xi) + \omega \Sigma_{count}(\varphi,\Phi,g,\xi)  = \Sigma(\varphi_0,\Phi_0,g_0,\xi_{0}) + O(\omega^2) \;, 
\end{equation}
where $(\varphi_0, \Phi_0, g_0,\xi_{0},\zeta_{0})$ stand for  the bare fields, external sources, coupling constant and parameter $(\xi_0,\zeta_0)$:
\label{renormfs}
\begin{equation}
\varphi_{0}=Z^{1/2}_{\varphi}\,\varphi  \qquad\;,   \qquad
\Phi_{0}=Z_{\Phi}\,\Phi\,,  \qquad g_0 = Z_g g \,,  \qquad \xi_0 = Z_{\xi} \xi  \,, \qquad  \zeta_0 = Z_{\zeta} \zeta  \;, 
\end{equation}
where, for the renormalization factors $Z$ we write 
\begin{equation}
Z^{1/2}_{\varphi}=(1+\omega\,z_\varphi)^{1/2}=1+\omega \frac{z_{\varphi}}{2}+O(\omega^{2})\,,\qquad
Z_{\Phi}=1+\omega\,z_{\Phi}\,, \qquad Z_g = 1 +\omega z_g\,,    
\end{equation}
and 
\begin{equation}
 Z_\xi = 1 +\omega z_\xi \,, \qquad  Z_\zeta = 1 +\omega z_\zeta \;. \label{rn2} 
\end{equation}
In the present case, some  care is needed due to the natural mixing of quantities which have the same quantum numbers. In fact, by applying  the linearized Slavnov-Taylor operator  $\mathcal{B}_{\Sigma}$ to the local integrated polynomial $\Delta^{(-1)}$, eq.\eqref{fct}, we can easily notice that the field $\lambda^{a}$ and the quantity $\gamma_{5}\epsilon T^{a}$ have the same dimension and quantum numbers, see Appendix A. The same feature occurs  between  the $\mathfrak D^{a}$ field and the quantity $Y^{a}\gamma_{5}\epsilon$. As a consequence, taking into account the aforementioned mixing, for the renormalization of  $\lambda$ and $\mathfrak{D}$ we have: 
\begin{equation}
\label{lrenorm}
\lambda^{a\alpha}_{0}=Z^{1/2}_{\lambda}\,\lambda^{a\alpha}+\omega\, z_{1}\,T^{a}(\gamma_{5})^{\alpha\beta}\varepsilon_{\beta}
\end{equation}
and
\begin{equation}
\label{drenorm}
\mathfrak{D}^{a}_{0}=Z^{1/2}_{\mathfrak{D}}\,\mathfrak{D}^{a}+\omega\, z_{2}\,Y^{a\alpha}(\gamma_{5})_{\alpha\beta}\varepsilon^{\beta}\,. 
\end{equation}
More precisely, for the whole set of Z-factors, expressed in terms of the coefficients $a_{0}$, $a_{1}$, $a_{4}$, $a_{11}$, $a_{23}$, $a_{24}$, we have 
\begin{eqnarray}
Z_A^{1/2}&=&1+\omega\left(\frac{a_{0}}{2}+a_1\right)\,,\nonumber\\
Z_{g}&=&1-\omega\frac{a_{0}}{2}\,,\nonumber\\
Z_{\lambda}^{1/2}&=&1+\omega\left(\frac{a_{0}}{2}-a_{11}\right)\,,\nonumber\\
Z_{N}&=&(1+\omega a_{23})Z_{\lambda}^{-1}\,, \nonumber \\
Z_{\xi}&=&(1+\omega a_{4})Z_{g}^{-2}Z_{A}\,,\nonumber\\
Z_{\zeta} & = & (1+\omega (a_{24} -a_{23})) Z_{\lambda}^{-4} \;, 
\label{5}
\end{eqnarray}
while, the remaining renormalization factors are given by
\begin{eqnarray}
Z_T&=&Z_{g}^{-1/2}Z_{A}^{1/4}\nonumber\,,\\
Z_{\varepsilon}&=&Z_{g}^{1/2}Z_{A}^{-1/4}\nonumber\,,\\
Z_{Y}&=&Z_{g}^{-1/2}Z_{A}^{1/4}Z_{\lambda}^{-1/2}\nonumber\,,\\
Z_{b}^{1/2}&=&Z_{A}^{-1/2}\nonumber\,,\\
Z_{L}&=&Z_{A}^{1/2}\nonumber\,,\\
Z_{c}^{1/2}&=&Z_{\bar{c}}^{1/2}=Z_{K}=Z_{g}^{-1/2}Z_{A}^{-1/4}\nonumber\,,\\
Z_{\Lambda}&=&Z_{g}^{1/2}Z_{A}^{1/4}\nonumber\,,\\
Z_{J}&=&1\nonumber\,,\\
Z_{\mathfrak{D}}&=&1\nonumber\,,\\
Z_{X}&=&Z_{\lambda}^{-1/2}\nonumber\,,\\
Z_{\Omega}&=&Z_{A}^{-1/2}\nonumber\,,\\
Z_{\rho}&=&Z_{g}Z_{A}^{-1/2}\nonumber\,,\\
Z_{\tau}&=&Z_{g}^{1/2}Z_{A}^{-1/4}\nonumber\,,\\
Z_{j}&=&Z_{g}Z_{A}^{-1/2}\nonumber\,,\\
Z_{\chi}&=&Z_{g}\nonumber\,,\\
Z_{R}&=&Z_{g}^{-1/2}Z_{A}^{1/4}Z_{\lambda}^{-1}.
\label{3}
\end{eqnarray}
with
\begin{equation}
z_1=-z_2=a_{11}-\frac{a_0}{2}\,.
\label{4}
\end{equation}
This ends the all order algebraic renormalization proof of the supersymmetric Yang-Mills theories  $\mathcal{N}=1$ in presence of the local composite operators $A^2$, $A_\mu \gamma_\mu \lambda$ and $\bar{\lambda}\lambda$. \\\\Before concluding this section, a few remarks are in order. From the renormalization factor $Z_{j}=Z_{g}Z_{A}^{-1/2}$ of the source related to the composite operator $A^2$, we observe that the non-renormalization theorem expressed by the equation \eqref{nrAA} remains valid as well as the non-renormalization of the ghost-antighost-gluon vertex, as expressed by $Z_g Z_c Z_A^{1/2}=1$. Nevertheless, as one can observe form the renormalization factor $Z_N$, which depends from the free parameter $a_{23}$, it does not turn out possible to express the anomalous dimension of the gluino operator $ \bar{\lambda}^{a\alpha}\lambda^{a}_{\alpha}$ in a similar way to that of $A^2$. To some extent, this can be understood as a consequence of the use of the  Wess-Zumino gauge in which supersymmetry and gauge transformations are put together, giving rise to a supersymmetry  algebra which does not close on space-time translations. The same reasoning applies here also to the renormalization factors $Z_A$ and $Z_\lambda$ which turns out in fact to be different from each other, see \cite{Capri:2014jqa} for an explicit higher loop calculations of these  factors. 

\section{Conclusion}

In this work, the $\mathcal{N}=1$ supersymmetric Yang-Mills theory in the presence of the composite local operators $A^2$, $A_\mu\gamma_\mu \lambda$ and $ \bar{\lambda}\lambda$ has been analysed. \\\\Adopting the Wess-Zumino gauge and the Landau gauge-fixing condition, all the above operators have been included in the starting action by means of the construction of a generalized BRST operator $Q$ \cite{White:1992ai,Maggiore:1994dw,Maggiore:1994xw,Maggiore:1995gr,Maggiore:1996gg,Ulker:2001rc,
Capri:2014jqa}, encoding both gauge and supersymmetry transformations, eq.\eqref{Q}. The operator $Q$ turns out to be nilpotent when acting on local integrated polynomials in the fields and their derivatives, eq.\eqref{Q2}. As a consequence, the resulting action $\Sigma$, eq.\eqref{SSYM}, is $Q$-invariant. \\\\Further, using the algebraic renormalization procedure \cite{Piguet:1995er}, an all order proof of the renormalizability of $\Sigma$ has been achieved. \\\\As already mentioned, the present analysis can be considered as a first step towards a possible understanding of the formation of the dimension two condensate $\langle A^2 \rangle$ in $\mathcal{N}=1$ Super Yang-Mills and of its eventual relationship with the well established condensate $ \langle \bar{\lambda}\lambda \rangle $ \cite{Veneziano:1982ah} as well as with the confining character of $\mathcal{N}=1$ pure Super Yang-Mills. It would be certainly worth to investigate if a non-vanishing  $\langle A^2 \rangle$ could be, somehow, accommodated together with $ \langle \bar{\lambda}\lambda \rangle $ to keep a vanishing vacuum energy, as required by the supersymmetric nature of the theory. An indication that such a possibility might be realized can be taken from  \cite{Capri:2014xea}, where an attempt to generalize  the Gribov-Zwanziger confinement mechanism has been investigated for $\mathcal{N}=1$ Super Yang-Mills. Any progress in this direction will be reported soon.

\section*{Acknowledgments}

The Conselho Nacional de Desenvolvimento Cient\'{\i}fico e
Tecnol\'{o}gico (CNPq-Brazil), the SR2-UERJ, the
Coordena{\c{c}}{\~{a}}o de Aperfei{\c{c}}oamento de Pessoal de
N{\'{\i}}vel Superior (CAPES)  are gratefully acknowledged.

\newpage
\begin{appendix}
\section{Tables of Quantum numbers}
We display here the  quantum numbers of all fields, sources and parameters of the model. In the following we shall employ the notation ``C'' for denoting the commuting nature of fields, sources and parameters; and ``A'' in the anti-commuting case

\begin{table}[htp]\centering
\begin{tabular}{|c|c|c|c|c|c|c|}
\hline
Fields & $A_{\mu}^{a}$ & $b^{a}$ & $c^{a}$ & $\bar{c}^{a}$ & $\lambda^{a\alpha}$ & $\mathfrak D^{a}$ \\ \hline
dim & $1$ & $2$ & $1$ & $1$ & $\phantom{\Big|}\!\frac{3}{2}\phantom{\Big|}\!$ & $2$ \\ \hline
gh  & $0$ & $0$ & $1$ & $-1$ & $0$ & $0$ \\ \hline
$Nature$ & $C$ & $C$ &  $A$ & $A$ & $A$ & $C$  \\ \hline
\end{tabular}
\end{table}
\begin{table}[htp]\centering
\begin{tabular}{|c|c|c|c|c|c|c|c|c|c|c|c|c|c|c|}
\hline
Sources & ${\Omega}^{a}_{\mu}$ & $K^{a}_{\mu}$ & $\Lambda^{a\alpha}$ & $T^{a}$ & $J^{a}$ & $L^{a}$ & $Y^{a\alpha}$ & $X^{a\alpha}$& $j$ & $\chi$ & $\rho^{\alpha}_{\mu}$ & $\tau^{\alpha}_{\mu}$ & $R$ & $N$ \\ \hline
dim  & $3$ & $2$ & $3$ & $1$ & $2$ & $2$ & $\phantom{\Big|}\!\frac{3}{2}\phantom{\Big|}\!$ & $\phantom{\Big|}\!\frac{5}{2}\phantom{\Big|}\!$& $2$ & $1$ & $\phantom{\Big|}\!\frac{3}{2}\phantom{\Big|}\!$ & $\phantom{\Big|}\!\frac{1}{2}\phantom{\Big|}\!$& $0$ & $1$ \\ \hline
gh  & $0$ & $-1$ & $-1$ & $-1$ & $0$ & $-2$ & $-1$ & $0$& $0$ & $-1$ & $0$ & $-1$& $-1$ & $0$\\ \hline
$Nature$ & $C$ & $A$ & $A$ & $A$ & $C$ & $C$ & $C$ & $A$& $C$ & $A$ &  $A$ & $C$&  $A$ & $C$  \\ \hline
\end{tabular}
\end{table}
\begin{table}[htp]\centering
\begin{tabular}{|c|c|c|c|c|c|}
\hline
Parameters & $\epsilon^{\alpha}$ & $\bar{\epsilon}^{\alpha}$& $\xi$& $\zeta$\\ \hline
dim & $\phantom{\Big|}\!\frac{1}{2}\phantom{\Big|}\!$& $\phantom{\Big|}\!\frac{1}{2}\phantom{\Big|}\!$& $0$& $0$\\ \hline
gh  & $1$ & $1$& $0$& $0$\\ \hline
$Nature$ & $C$ & $C$& $C$& $C$\\ \hline
\end{tabular}
\end{table}

\section{Notations and conventions in Euclidean space-time}
\label{notations}
\noindent
\textbf{Units}: $\hbar=c=1$.

\noindent
\textbf{Euclidean metric}: $\delta_{\mu\nu}=diag(+,+,+,+)$.

\noindent
\textbf{Wick rotations:} $X_0\rightarrow -iX_4\Rightarrow\partial_0\rightarrow+i\partial_4$, $A_0\rightarrow+iA_4$.

\noindent
\textbf{Gamma matrices:}
\begin{equation}
\gamma_4=\left( \begin{array}{cc}
0 & \mathbb{I} \\
\mathbb{I} & 0 \end{array} \right),\quad
\gamma_k=-i\left( \begin{array}{cc} 0 & \sigma_k \\ -\sigma_k & 0 \end{array} \right),\quad
\mathbb{I}=\left( \begin{array}{cc}
1 & 0 \\
0 & 1 \end{array} \right),\quad \mathrm{with}\,\,k=1,2,3\,.
\end{equation}

\noindent
\textbf{Pauli matrices}:
\begin{equation}
\sigma_4\equiv\mathbb{I}=\left( \begin{array}{cc}
1 & 0 \\
0 & 1 \end{array} \right),~~
\sigma_1=\left( \begin{array}{cc} 0 & 1\\ 1 & 0 \end{array} \right),\quad
\sigma_2=\left( \begin{array}{cc} 0 & -i\\ i & 0 \end{array} \right),\quad
\sigma_3=\left( \begin{array}{cc} 1 & 0 \\ 0 & -1 \end{array} \right).
\end{equation}

\noindent
The Gamma matrices obey the following properties:
\begin{eqnarray}
 \gamma_\mu=\gamma_\mu^\dagger\\
  \{\gamma_\mu,\gamma_\nu\}&=&2\delta_{\mu\nu}
\end{eqnarray}

\noindent
We also define the $\gamma_5$ matrix as:
\[\gamma_5=\gamma_4\gamma_1\gamma_2\gamma_3=\left(\begin{array}{cc}
                                                   \mathbb{I} & 0\\
                                                   0 & -\mathbb{I}\\
                                                  \end{array}\right)\]

\noindent
with the following properties:
\begin{equation}
 \{\gamma_5,\gamma_\mu\}=0,~~(\gamma_5)^2=\mathbb{I},~~\gamma_5^\dagger=\gamma_5
\end{equation}

\noindent
The charge conjugation matrix is:
\begin{equation}
\label{Cmtrx}
\mathcal{C}=\gamma_4\gamma_2=i\left(\begin{array}{cc} \sigma_2 & 0 \\ 0 & -\sigma_2\\ \end{array}\right)
\end{equation}

\noindent
with the following properties:
\begin{equation}
 \mathcal{C}^{-1}=-\mathcal{C}=\mathcal{C}^T,~~\mathcal{C}^{-1}\gamma_\mu\mathcal{C}=-
\gamma_\mu^T
\end{equation}

\noindent
The $\sigma^{\mu\nu}$ tensor is defined as
\begin{equation}
  (\sigma_{\mu\nu})^{~\beta}_\alpha\equiv\frac{1}{2}[\gamma_\mu,\gamma_\nu]^{~\beta}_\alpha
\end{equation}
and has the property $\sigma_{\mu\nu}^\dagger=-\sigma_{\mu\nu}$.

\noindent
\textbf{Majorana fermions:}

\noindent
The Majorana condition reads:
\begin{equation}
\label{mjconj}
 \lambda^\mathcal{C}=\lambda=\mathcal{C}\bar{\lambda}^T~~
\Longleftrightarrow~~\bar{\lambda}=\lambda^T\mathcal{C}\;,
\end{equation}
leading to the following relations
\begin{equation}
\bar{\lambda}\gamma_{\mu}\epsilon = \bar{\epsilon}\gamma_{\mu}\lambda \qquad \text{and} \qquad \bar{\lambda}\gamma_{\mu}\gamma_{5}\epsilon = - \bar{\epsilon}\gamma_{\mu}\gamma_{5}\lambda\;.
\end{equation}

\noindent
\textbf{Fierz identity (in Euclidean space-time):}
\begin{eqnarray}
\epsilon_{1}\bar{\epsilon}_{2} &=&  \frac{1}{4}(\bar{\epsilon}_{2}\epsilon_{1})\mathbb{I} 
+ \frac{1}{4}(\bar{\epsilon}_{2}\gamma_{5}\epsilon_{1})\gamma_{5}
+ \frac{1}{4}(\bar{\epsilon}_{2}\gamma_{\mu}\epsilon_{1})\gamma_{\mu}
- \frac{1}{4}(\bar{\epsilon}_{2}\gamma_{\mu}\gamma_{5}\epsilon_{1})\gamma_{\mu}\gamma_{5}\nonumber \\
&&
- \frac{1}{8}(\bar{\epsilon}_{2}\sigma_{\mu\nu}\epsilon_{1})\sigma_{\mu\nu}    \;.
\end{eqnarray}

\noindent {\bf Indices notations}:
\begin{center}
\begin{tabular}{ll}
$\bullet$&The Lorentz indices: $\mu,\nu,\rho,\sigma,\lambda\in\{1,2,3,4\}$\,;$\phantom{\Bigl|}$\\
$\bullet$&The Spinor indices: $\alpha,\beta,\gamma,\delta,\eta\in\{1,2,3,4\}$\,;$\phantom{\Bigl|}$\\
$\bullet$&The $SU(N)$ group indices: $a,b,c,d,e\in\{1,\dots,N^{2}-1\}$\,;$\phantom{\Bigl|}$\\
\end{tabular}
\end{center}

\end{appendix}

\end{document}